\documentclass{svjour3}                     % onecolumn (standard format)
\smartqed  % flush right qed marks, e.g. at end of proof
\usepackage{graphicx}
\journalname{Few-Body Systems (FB20)}
\begin{document}

\title{
%Insert your title here
Search of Kaonic Nuclear States at the SuperB factory
\thanks{Presented at the 20th International IUPAP Conference on Few-Body Problems in Physics, 20 - 25 August, 2012, Fukuoka, Japan}
}
\subtitle{
%Do you have a subtitle?\\ If so, write it here
%Subtitle
}

%\titlerunning{Short form of title}        % if too long for running head

\author{S. Marcello  \and F. De Mori \and A. Filippi
         %etc.
}

%\authorrunning{Short form of author list} % if too long for running head

\institute{S. Marcello \at
              Torino University, Dept. of Physics and INFN Torino, via P. Giuria 1, Torino 10125, Italy \\
              Tel.: +39-011-6707321\\
              Fax: +39-011-6707324\\
              \email{marcello@ph.unito.it}           %  \\
%             \emph{Present address:} of F. Author  %  if needed
           \and F. De Mori \at
           Torino University, Dept. of Physics and INFN Torino,
           via P. Giuria 1, Torino 10125, Italy\\
           \and A. Filippi \at
           INFN Torino, via P. Giuria 1, Torino 10125, Italy
}

\date{Received: date / Accepted: date}
% The correct dates will be entered by the editor

\maketitle

\begin{abstract}
The search of nuclear bound states of $\bar{K}$ in few-body nuclear systems
such as $K^-pp$, can be extended from the nuclear medium to the vacuum,
using the glue-rich $\Upsilon(1S)$ decays at B-factories.    
Here the possibility for such a measurement at the future 
SuperB factory is discussed.

%Insert your abstract here. Include keywords, PACS and mathematical
%subject classification numbers as needed.
% \subclass{MSC code1 \and MSC code2 \and more}
\end{abstract}

\section{Introduction}
\label{intro}

The existence of $\bar{K}$-nuclear bound states, such as $K^-pp$, $K^-ppn$ 
and $K^-ppnn$ has been predicted by Akaishi and Yamazaki \cite{DBKS}. 
They constructed a phenomenological $\bar{K}-N$ potential, 
using experimental data on free $\bar{K}-N$ scattering lengths
and on kaonic hydrogen X-rays, with the peculiar ansatz
that the $\Lambda(1405)$ resonance is a $\bar{K}-N$ bound state, 
instead of a simple baryon with a three-quark structure.
In this model the strongly attractive $\bar{K}-N$ interaction, in the 
isospin $I = 0$ state, may lead to tremendously 
condensed systems with a cluster structure and a central nucleon density 
as high as 1.5 fm$^{-3}$. In this model very narrow ($<$ 100 MeV) 
and deeply bound ($\sim100$ MeV) kaon states (DBKS) are expected. 

The debate on their existence is still open and different theoretical 
approaches, quite in disagreement, have been developed (see \cite{Filippi}
for a summary of theoretical issues). 
The main question concerns whether the potential is shallow or deep 
and, then, if the decay width is narrow enough, compared to the binding energy,
to allow their observation. 
Furthermore, since Final State Interactions (FSI) could fake a signal, 
positive experimental results could be attributed to such
an effect.

Since DBKS are very dense systems, made of a strange particle 
and a few nucleons, the study of this system 
is related to kaon condensation \cite{Kaplan-KCond,Brown-KCond} and 
is important to understand 
astrophysical objects such as the neutron stars, where the nuclear density
could reach ten times the ordinary one and where strangeness degrees of freedom
could play a crucial role \cite{Weber}. This could lead to the formation 
of strange quark matter, which could be more stable than the ordinary 
nuclear matter. 
For this reason, even after some years, the study of DBKS keeps on being 
a hot topic in hadron physics.

 The first observation of a few hints for the lightest nuclear DBKS, 
the $K^-pp$,
was done by the FINUDA experiment \cite{FINUDA} at the DA$\Phi$NE $e^+e^-$ 
collider in Frascati, stopping  $K^-$ of low energy in very thin targets 
of $p$-shell nuclei. Here, a narrow peak ($\Gamma \sim 70$ MeV) has been 
observed in the $\Lambda$p invariant mass spectrum 
at a mass $\sim$2.26 GeV/c$^2$ corresponding to a binding energy 
B $\sim 115$ MeV. 
Therefore, the $\Lambda$p pairs,
which exhibit a clean back-to-back topology,  
have been interpreted as the decay products of such a state.

New analyses of old experiments, such as OBELIX \cite{Obelix} at CERN 
with $\bar{p}-^4He$ annihilations at rest
and DISTO \cite{DISTO} at SATURNE with the exclusive 
$pp \rightarrow p K^+ \Lambda$ production at $T_p$ = 2.85 GeV, 
claim the observation of DBKS. 
Nevertheless, their existence has not been established yet 
and further measurements 
are planned in the near future at several facilities. Since all of them
uses kaon beams interacting with nuclear targets or heavy ion 
collisions, where the formation of the nuclear kaonic state occurs 
in the nuclear medium, new experiments in different environments 
could help to find out signatures of the existence of such states. 
In particular, a search of production of DBKS in the vacuum, 
rather than in the medium, can be performed at B-factories,  
in the direct decays of the lightest bottomonium.

\section{Production of baryons and deuteron 
in $\Upsilon(1S)$ decays}
\label{sec:1}
In the bottomonium spectrum below the $B\bar{B}$ threshold 
there are three states, 
the $\Upsilon(1S)$, $\Upsilon(2S)$ and $\Upsilon(3S)$. 
The large mass of these resonances ($\sim 10$ GeV/c$^2$) is suitable to produce 
baryons and anti-baryons, and few-body bound systems, as well.
It should be noted that the branching ratio for the strong decays 
of $\Upsilon(1S)$ amounts to $\sim80\%$. 
These direct decays proceed via three gluons, hence the decay
products are the result of their hadronization.
These glue-rich decays might also produce exotic multiquark states. 
For the $\Upsilon(2S)$ and $\Upsilon(3S)$ resonances, direct radiative decays,
both e.m. and hadronic, compete with the gluon annihilation modes;
but these states can be used to produce the $\Upsilon(1S)$.

It is well known, since many years, that baryon production from $\Upsilon$(1S) 
is enhanced with respect to the continuum hadronization \cite{Behrends}. 
In particular, the CLEO experiment \cite{BaryonCLEO} at CESR, 
studying the inclusive 
production of baryons/antibaryons in $e^+e^-$ collisions 
at $\sqrt{s} \sim 10$ GeV, 
finds that the enhancements of per-event total particle yields 
for $\Lambda$ hyperon, proton and antiproton are about a factor of two  
in the $\Upsilon (1S) \rightarrow ggg$ decays as compared to the 
non resonant nearby $q\bar{q}$ continuum ($e^+e^- \rightarrow q\bar{q}$).
The enhancement for $\Lambda$ production is the highest one ($\sim 2,7$). 
These results are not accounted for by the JETSET 7.3 fragmentation model. 
For the decays through gg$\gamma$ the enhancements 
are smaller (except for $\Lambda$'s which is a factor $\sim$ 2) 
and more in agreement with JETSET expectations.

Most interesting, concerning the differences between
quark/gluon-fragmentation in the $\sqrt{s} \sim 10$ GeV energy range, 
is the experimental evidence of bound state production, 
such as deuteron ($d$) and antideuteron ($\bar{d}$),
in gluonic decays of $\Upsilon$(1S) and $\Upsilon$(2S), 
that was first observed by ARGUS \cite{ARGUS} with very low statistics and, 
recently, by CLEO \cite{AntiD-CLEO}.
In particular, CLEO finds that the branching ratio 
$\cal{B}$($\Upsilon (1S) \rightarrow \bar{d}X$)
is of the order of 3$\times$10$^{-5}$. 
Therefore, the results show
that $\bar{d}$ production is at least three times more likely in 
the hadronization of $\Upsilon (nS)$ (for n = 1, 2) through ggg and gg$\gamma$, 
compared to the hadronization of $q\bar{q}$ for the continuum production, 
where only an upper limit of 1$\times$10$^{-5}$ 
has been evaluated from the 
$e^+e^- \rightarrow \bar{d} X$ cross section. 
%where no significant production has been has been observed.
Besides, an upper limit of 1$\times$10$^{-5}$ has been 
measured for the production of $\bar{d}$ from $\Upsilon(4S)$.
 
CLEO also investigated how often the $\bar{d}$ baryon number 
is compensated by a $d$ or by a combination of two nucleons (n, p): 
the first case occurs 1\% of the times.  
Since the theoretical description of $d$ or $\bar{d}$ formation is based 
on the statistical coalescence model \cite{Coalescence}, where a $\bar{n}$ 
and $\bar{p}$ close to each other in phase 
space bind together, a precise determination of such a percentage is important,
because double coalescence is unlikely and a different production 
mechanism, such as a primary (globally) production might be involved.

It can be also interesting to note that in heavy ion collisions the 
formation of bound systems, 
such as light nuclei/anti-nuclei or hypernuclei/anti-hypernuclei, 
occurs through coalescence processes. For instance, 
the STAR \cite{STAR-antihyp} 
experiment at RHIC has measured the ratios of the yields of 
anti-$^3_{\Lambda}H$ and $^3_{\Lambda}H$ and of anti-$^3He$ and $^3He$ 
in Au-Au collisions
at $\sqrt{s_{NN}}$ = 200 GeV, which amount to $\sim$ 0.5,
favouring the coalescence hypothesis. At STAR (anti)hypernuclei are produced
with similar yields of (anti)nuclei, different from what happens 
at lower energies.
Recently, the hunt for DBKS is also started   
in heavy ion collision experiments \cite{FOPI}, 
but their identification in the medium is a challenge.

\section{Search of kaonic nuclear states at the SuperB factory}
\label{sec:2}
Since bound state systems, such as antideuterons, have been 
found in the decays of $\Upsilon(nS)$, with n = 1, 2, 
we propose to search for DBKS at the future SuperB factory \cite{SUPERB}, 
which will be built in the Tor Vergata University Campus in Rome. 
Here, large numbers of heavy leptons and heavy quark mesons 
will be produced using an e$^+$e$^-$ asymmetric collider, operated at 
a c.m. energy corresponding to the rest mass of the bottomonium 
resonance $\Upsilon(4S)$ with a luminosity of 10$^{36}$cm$^{-2}$s$^{-1}$, 
looking for new physics.
Thanks to the nano-beam scheme, at LNF-INFN, an unprecedented 
integrated luminosity of 10 $ab^{-1}$/year will be achieved.  
A magnetic spectrometer will be installed at the machine,
covering a large solid angle and designed to detect 
and fully reconstruct both charged and neutral particles with high efficiency 
and energy resolution.
Operating the machine at an energy below the $B\bar{B}$ 
threshold it will be possible to select the rest mass of 
$\Upsilon$(nS) (with n = 1, 2, 3) in order to study the production of DBKS
in the decays which proceed through the hadronization 
of three gluons. 

The lightest nuclear DBKS, the  $K^-$pp, can be identified through 
its $\Lambda p$
decay mode, searching for a narrow peak ($<$ 100 MeV/c$^2$) at a mass 
of about 2.25 GeV/c$^2$ 
in the $\Lambda p$ invariant mass spectrum and 
measuring the $\Lambda-p$ angular correlations, which can give imporant 
hints on the nature of the event. 
We can rely on the high performance of the SuperB apparatus 
for particle ID and full topological event reconstruction for an effective
identification of the DBKS.

Since there is no medium, FSI are negligible with respect to 
experiments using kaon beams on nuclear targets or heavy ion 
collisions. Moreover, we expect that the identification from the vacuum
is easier and cleaner than in heavy ion collisions.

\section{Conclusions}
\label{sec:3}
Nowadays, the study of kaonic nuclear states is a hot topic in hadron physics,
because their existence is related to kaon condensation and to the physics of 
the core of neutron stars.
Theoretical models disagree about the possibility to observe such states
and only scarce experimental measurements in the nuclear medium are available.
More and new experiments in different environments would be useful for a 
comparative analysis of the results.  
Search of the light $K^-pp$ state could be extended from the nuclear medium
to the vacuum, looking for its production in the strong decays 
of $\Upsilon(1S)$ at future B-factories, 
taking advantage of the high luminosity of these machines.

Measurements of the production yields of this state from the 
lightest bottomonium 
should give important clues, not only about the nature of these 
states, but also 
about the hadronization processes of quarks and gluons.

% Non-BibTeX users please use

\end{document}